# Identification of Physical Properties in Acoustic Tubes Using Physics-Informed Neural Networks

Kazuya YOKOTA*, Masataka OGURA** and Masajiro ABE***
* Department of Mechanical Engineering, Nagaoka University of Technology
1603-1 Kamitomioka, Nagaoka, Niigata 940-2188, Japan
E-mail: yokokazu@vos.nagaokaut.ac.jp
**Center for Integrated Technology Support, Nagaoka University of Technology, Japan
***Department of System Safety Engineering, Nagaoka University of Technology, Japan

**Abstract**
Physics-informed Neural Networks (PINNs) is a method for numerical simulation that incorporates a loss function corresponding to the governing equations into a neural network. While PINNs have been explored for their utility in inverse analysis, their application in acoustic analysis remains limited. This study presents a method to identify loss parameters in acoustic tubes using PINNs. We categorized the loss parameters into two groups: one dependent on the tube's diameter and another constant, independent of it. The latter were set as the trainable parameters of the neural network. The problem of identifying the loss parameter was formulated as an optimization problem, with the physical properties being determined through this process. The neural network architecture employed was based on our previously proposed ResoNet, which is designed for analyzing acoustic resonance. The efficacy of the proposed method is assessed through both forward and inverse analysis, specifically through the identification of loss parameters. The findings demonstrate that it is feasible to accurately identify parameters that significantly impact the sound field under analysis. By merely altering the governing equations in the loss function, this method could be adapted to various sound fields, suggesting its potential for broad application.

***Keywords*** : Physics-informed Neural Networks (PINNs), Acoustic analysis, Acoustic tube, Inverse analysis, Wave equation

## 1. Introduction

The physics-informed neural network (PINN) is a machine learning method proposed to address forward and inverse analyses on partial differential equations (Raissi et al., 2019). Unlike traditional machine-learning numerical simulations that necessitate pre-labeled data, PINNs integrate a loss function corresponding to the governing equations into the neural network, transforming the challenge of solving partial differential equations into a task of minimizing these loss functions. The required derivative calculations for the loss function are facilitated by the automatic differentiation capabilities of the neural network (Schmidhuber, 2015). This unique architecture enables PINNs to conduct mesh-free forward and inverse analyses without prior data preparation. Recently, PINNs have achieved significant advancements in inverse analysis across various fields, including nanoengineering (Chen et al., 2020), flow field estimation from sparse data (Cai et al., 2021; Mai et al., 2024), and blood circulation analysis (Jeong et al., 2024).

In the realms of acoustics and vibration, while constructing PINNs for the wave equation presents challenges due to complexities in reflection, diffraction, and dynamics of amplitude and frequency (Moseley et al., 2020), applications in seismic wave analysis have been reported (Karimpouli and Tahmasebi, 2020; Zhang et al., 2023). PINNs are also being explored in audible-range acoustic problems, such as acoustic holography (Olivieri et al., 2021; Kafri et al., 2023), sound field reconstruction (Silva Garzon et al., 2023), and impulse response estimation in acoustic fields (Karakonstantis et al., 2024).

However, reports on PINNs applications in acoustic tubes remain scant. The modeling of sound fields in acoustic tubes has broad applications, including the design of automobile mufflers (Kashikar et al., 2021), active noise control in duct (Bai and Zeung, 2002), vocal tract analysis (Yokota et al., 2019), and brass/wind instrument design (Tournemenne et al., 2017). Addressing this gap, we previously developed ResoNet, a PINNs designed for one-dimensional (1D) resonance analysis in acoustic tubes (Yokota et al., 2023). ResoNet effectively analyzes acoustic resonance in the time domain by incorporating a loss function for periodic solutions and has demonstrated high accuracy in both forward and inverse analyses.

This study explores a method for identifying loss parameters in 1D acoustic tubes using PINNs, aiming to broaden the application of PINNs to diverse inverse problems associated with acoustic tubes. Employing ResoNet, the sound for parameter identification is emitted at one endpoint of the tube, with sound pressure measurements acquired by a microphone at the other endpoint to identify loss parameters from the data. This study presents two primary contributions: first, it presents a practical formulation for the problem of identifying loss parameters in acoustic tubes using PINNs; second, it discusses the potential and limitations of using PINNs for property identification in time-domain acoustic analysis.

The remainder of this paper is structured as follows: Section 2 outlines the governing equations and formulates the problem of loss parameter identification; Section 3 details the structure of PINNs for this task; Section 4 presents the results of parameter identification and their validation; Section 5 concludes the paper and discusses the future applicability of PINNs to the inverse problem for acoustic tubes.

## 2. Governing equations and problem formulation

This section outlines the governing equations for 1D sound wave propagation in an acoustic tube and the formulation of the problem for identifying loss parameters.

### 2.1 Governing equations for 1D acoustic propagation

We examine the propagation of 1D sound waves in an acoustic tube with a variable cross-sectional area, shown in Fig. 1. Here, $x$ represents the axial position along the tube, and $A(x)$ denotes the cross-sectional area at a given position. The variables $p$ and $U$ signify the sound pressure and the volume velocity within the tube, respectively. The mathematical expression for 1D sound wave propagation in such a tube is presented as follows (Flanagan, 2013):

$$\frac{dU}{dx} = -Gp - \frac{A}{K}\frac{\partial p}{\partial t}, \tag{1}$$

$$\frac{dp}{dx} = -RU - \frac{\rho}{A}\frac{\partial U}{\partial t}, \tag{2}$$

where $G$ is an energy loss parameter related to heat conduction at the wall surface, $R$ is an energy loss parameter related to the viscosity of the wall surface of the acoustic tube, $\rho$ is the air density, and $K$ is the bulk modulus. $G$ and $R$ are calculated using the following theoretical equations for a rigid wall surface with infinite thermal conductivity.

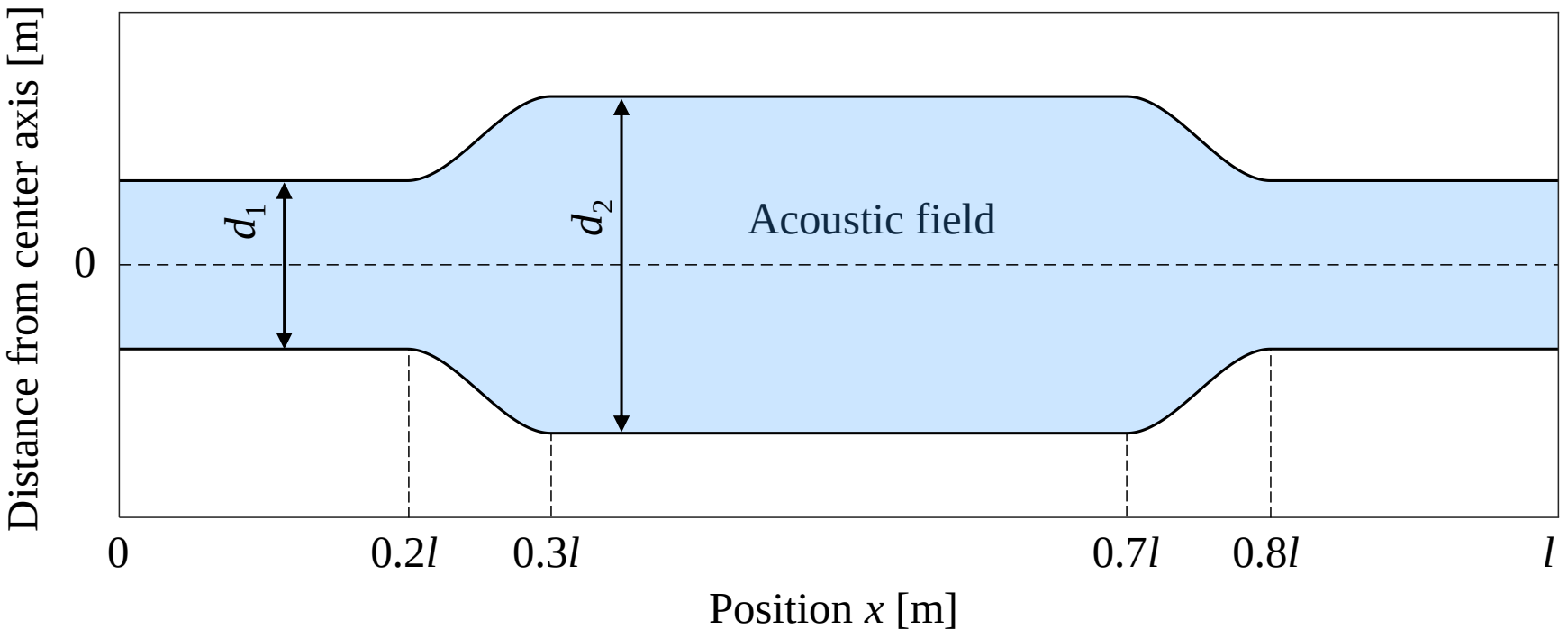

Fig. 1 Shape of an acoustic tube analyzed in this study. The model of an acoustic tube with a variable cross-sectional area has two sections, one with a diameter of $d_1$ and the other with a diameter of $d_2$. The shape of the part with changing diameter is interpolated by a Piecewise Cubic Hermite Interpolating Polynomial.

$$G = S\frac{\eta - 1}{\rho c^2}\sqrt{\frac{\lambda\omega_c}{2c_p\rho}}, \tag{3}$$

$$R = \frac{S}{A}\sqrt{\frac{\omega_c\rho\mu}{2}}, \tag{4}$$

where $S$ is the circumference of the acoustic tube, $\eta$ is the ratio of the specific heat of air at constant pressure to the specific heat of air at constant volume, $c$ is the speed of sound in air, $\lambda$ is the thermal conductivity of air, $\omega_c$ is the angular frequency used to calculate the loss, $c_p$ is the specific heat of air at constant pressure, and $\mu$ is the viscosity coefficient of air. Note that $G$ and $R$ are also functions of $x$, since $A$ and $S$ are functions of $x$ here.

## 2.2 Boundary conditions

At $x = 0$, where sound is emitted for parameter identification, we assume that the flow velocity waveform is known, allowing us to apply a forced flow boundary condition at $x = 0$. The waveform used is the Rosenberg wave (Rosenberg, 1971), as illustrated in Fig. 2. The fundamental frequency of this waveform is 261.6 Hz, corresponding to C4 on the musical scale, and it is processed through a low-pass filter with a cutoff frequency of 2 kHz.

At $x = l$, the open end of the tube, radiation boundary conditions are applied. Here, $U_l$and $p_l$ represent the volume velocity and sound pressure at $x = l$, respectively. The relationship between these variables is described by the following equation (Ishizaka and Flanagan, 1972):

$$(U_l - U_r)R_r = L_r\frac{dU_r}{dt}, \tag{5}$$

$$p_l = (U_l - U_r)R_r, \tag{6}$$

where $U_r$ is the volume velocity used to calculate radiation, and $R_r$ and $L_r$ are approximated using the following equations when the aperture end is surrounded by an infinitely flat baffle.

$$R_r = \frac{128\rho c}{9\pi^2 A_l}, \tag{7}$$

$$L_r = \frac{8\rho}{3\pi\sqrt{\pi A_l}}, \tag{8}$$

where $A_l$ is the cross-sectional area at $x = l$.

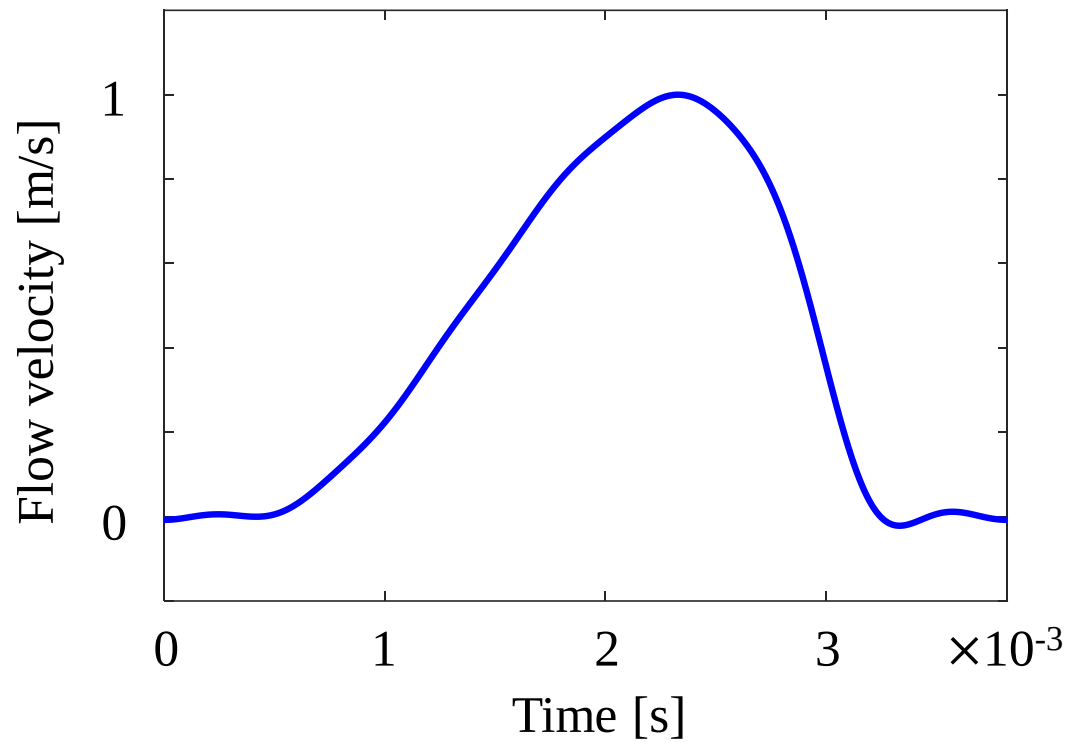

Fig. 2 Rosenberg waveform as a boundary condition at $x = 0$. The Rosenberg wave exhibits overtone components over a wide frequency range, and since the overtone components are monotonically decreasing with frequency, this waveform can be directly emitted by a loudspeaker.

### 2.3 Problem formulation

The methodology for identifying loss parameters in this study is shown in Fig. 3. An acoustic output positioned at $x = 0$ induces resonance within the tube, which is subsequently captured by a microphone located at $x = l$. The identification process for the loss parameters $G$ and $R$ involves analyzing the measured particle velocity at $x = 0$ and the sound pressure at $x = l$ using PINNs.

## 3. Physics-informed Neural Networks for loss parameter identification

This section outlines the implementation of PINNs for acoustic analysis in acoustic tubes and describes the methodology for identifying loss parameters.

### 3.1 Structure of neural network

In this study, we analyze resonance in acoustic tubes using the PINNs structure shown in Fig. 4, where the inputs are $x$ and $t$ and the outputs are the sound pressure and volume velocity predictions $\hat{p}$ and $\hat{U}$. The configuration shown in Fig. 4 is derived from our previously proposed ResoNet (Yokota et al., 2023), a PINNs tailored for acoustic resonance analysis. While the original ResoNet model predicted velocity potentials, the neural network in the current study provides predictions of sound pressure and volume velocity. The lower part of the network outputs the prediction of $U_r$ in Eq. (5) and (6).

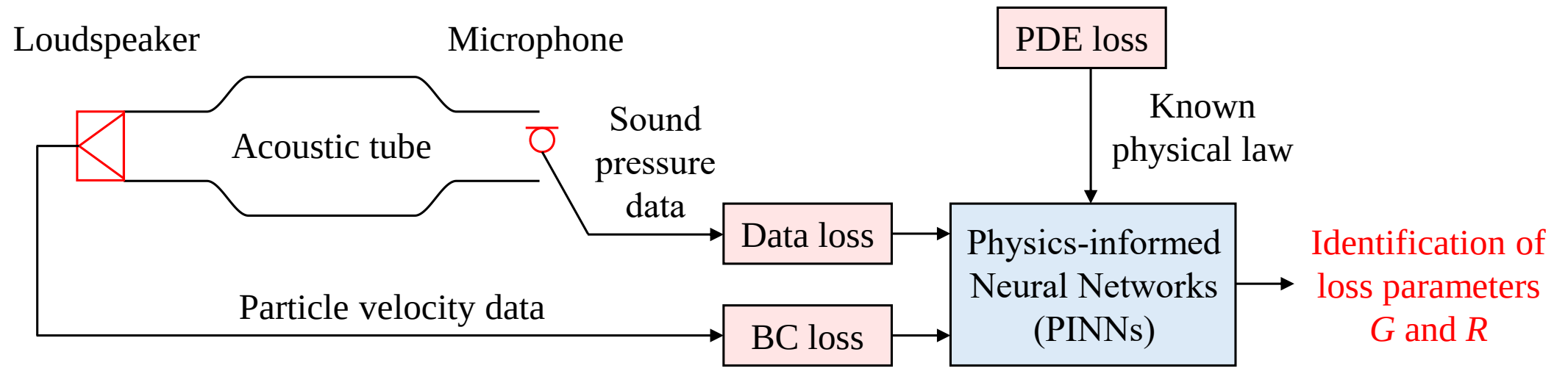

Fig. 3 Procedure for identifying the loss parameters. The loudspeaker-generated resonance is measured using a microphone at $x = l$. Based on the particle velocity waveform at $x = 0$ and the sound pressure waveform at $x = l$, loss parameters $G$ and $R$ in Eq. (1) and (2) are identified using PINNs.

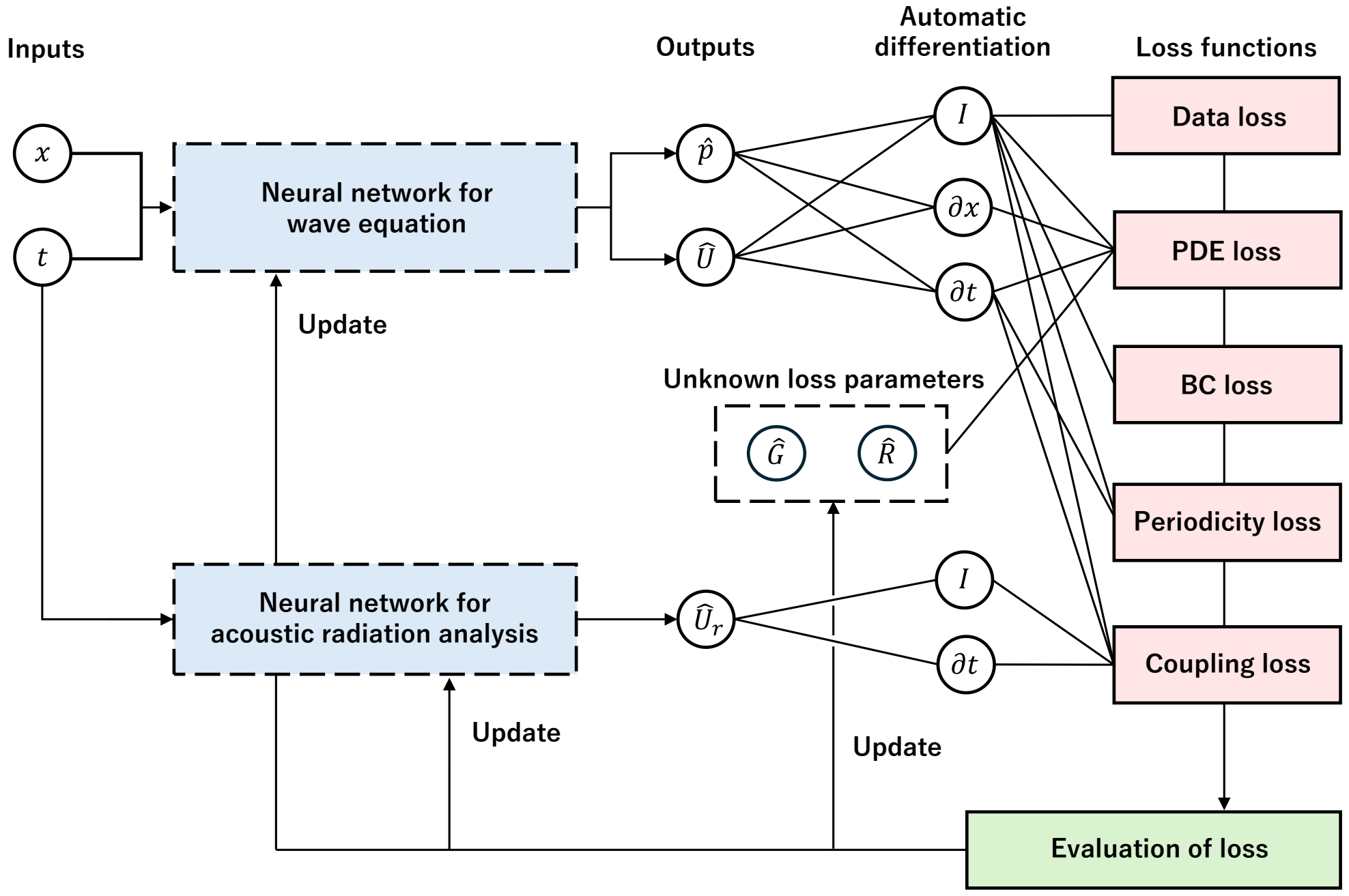

Fig. 4 PINN structure for loss parameter identification, which is based on our previously proposed ResoNet. The upper neural network outputs the solution of Eq. (1) and (2), while the lower network outputs the prediction of the volume velocity $\hat{U}_r$ with respect to the radiation evaluated using Eq. (5) and (6).

The upper and lower networks in Fig. 4 consist of fully-connected neural networks with residual connections, details of which can be found in our previous paper (Yokota, 2023). The activation function used in the neural network is snake (Ziyin et al., 2020), as shown in the following equation:

$$f(a) = a + \sin^2 a, \tag{9}$$

where $a$ is the input of the activation function.

Automatic differentiation is applied to the outputs $\hat{p}$, $\hat{U}$, and $\hat{U}_r$ to compute the four types of activation functions displayed on the right side of Fig. 4. Partial differential equation (PDE) loss $L_E$ compels the outputs $\hat{p}$ and $\hat{U}$ to conform to Eq. (1) and (2), which are defined as follows:

$$L_E = \frac{\lambda_{E1}}{N_E}\sum_{i=1}^{N_E}\left(\frac{d\hat{U}_i}{dx_i} + \hat{G}_i\hat{p}_i + \frac{A_i}{K}\frac{\partial\hat{p}_i}{\partial t_i}\right)^2 + \frac{\lambda_{E2}}{N_E}\sum_{i}^{N_E}\left(\frac{d\hat{p}_i}{dx_i} + \hat{R}_i\hat{U}_i + \frac{\rho}{A_i}\frac{\partial\hat{U}_i}{\partial t_i}\right)^2, \tag{10}$$

where $x_i \in [0, l]$, $t_i \in [0, T]$, $T$ is the simulation time (one resonance period in this study), $N_E$ is the number of collocation points for PDE loss, and $\lambda$ is the weight parameter. $\hat{G}$ and $\hat{R}$ in Eq. (10) are the estimated values of the loss parameters calculated by the identification method described in the following section.

The boundary condition at $x = 0$ is incorporated into the neural network through the boundary condition (BC) loss in Fig. 4. The BC loss $L_B$ for $x_i = 0$ is defined as follows:

$$L_B = \frac{\lambda_B}{N_B}\sum_{i=1}^{N_B}\left(\frac{\hat{U}_i}{A_i} - \bar{v}_i\right)^2, \tag{11}$$

where $N_B$ is the number of collocation points for BC loss, and $\bar{v}_i$ is the particle velocity data at $x = 0$, the boundary condition. $\bar{v}_i$ is Rosenberg wave, as shown in Fig. 2.

At $x = l$, the equations for acoustic wave propagation in Eqs. (1) and (2) are coupled with the equations for radiation in Eqs. (5) and (6). This relationship is incorporated into the neural network as the coupling loss $L_c$, which is defined for $x_i = l$ as follows:

$$L_C = \frac{\lambda_C}{N_C}\sum_{i=1}^{N_C}\left\{(\hat{U}_i - \hat{U}_{ri})R_r - L_r\frac{d\hat{U}_{ri}}{dx_i}\right\}^2 + \frac{\lambda_C}{N_C}\sum_{i=1}^{N_c}\left\{\hat{p}_i - (\hat{U}_i - \hat{U}_{ri})R_r\right\}^2, \tag{12}$$

where $N_C$ is the number of collocation points corresponding to the coupling loss.

As outlined in Section 2, parameter identification in this study is conducted using the microphone waveform in the resonance state. Consistent with our previously proposed ResoNet, only one period in the resonance state is analyzed. Let $\hat{p}_{i,P1}$ and $\hat{U}_{i,P1}$ represent the predicted values of sound pressure and volume velocity at $t = 0$, and let $\hat{p}_{i,P2}$ and $\hat{U}_{i,P2}$ represent those at $t = T$ ($T$: period). Since all physical quantities are equivalent at $t = 0$ and $t = T$ in the steady state, the periodicity loss $L_P$ ensures that the outputs of the neural network conform to a steady-state solution.

$$L_P = L_{P0} + L_{P1}, \tag{13}$$

where

$$L_{P0} = \frac{\lambda_{P0,U}}{N_P}\sum_{i=1}^{N_P}\left(\hat{U}_{i,P1} - \hat{U}_{i,P2}\right)^2 + \frac{\lambda_{P0,p}}{N_P}\sum_{i=1}^{N_P}\left(\hat{p}_{i,P1} - \hat{p}_{i,P2}\right)^2, \tag{14}$$

$$L_{P1} = \frac{\lambda_{P1,U}}{N_P}\sum_{i=1}^{N_P}\left(\frac{d\hat{U}_{i,P1}}{dt_i} - \frac{d\hat{U}_{i,P2}}{dt_i}\right)^2 + \frac{\lambda_{P1,p}}{N_P}\sum_{i=1}^{N_P}\left(\frac{d\hat{p}_{i,P1}}{dt_i} - \frac{d\hat{p}_{i,P2}}{dt_i}\right)^2, \tag{15}$$

where $N_P$ is the number of collocation points corresponding to the periodicity loss.

If the loss parameters $G$ and $R$ are known, forward analysis can be performed using PINNs. The loss function for the entire system $L_{forward}$ in the forward analysis is defined as follows:

$$L_{forward} = L_E + L_B + L_C + L_P. \tag{16}$$

Finally, the optimization problem of the neural network in the forward analysis is formulated as follows:

$$\min_{\theta} L_{forward}(\theta), \tag{17}$$

where $\theta$ is a trainable parameter of the neural network. By minimizing $L_{forward}$, the neural network can learn the governing equations and boundary conditions to perform numerical simulations. We optimize Eq. (17) through iterative calculations using Adam optimizer (Kingma, 2014) to determine $\theta$.

### 3.2 Identification method for the loss parameters $G$ and $R$

This section describes the parameter identification method when $G$ and $R$ are unknown. As expressed by Eq. (3) and (4), these loss parameters incorporate the cross-sectional area $A$ and circumference $S$, both of which are functions of $x$. Consequently, we distinguish the constants $G_c$ and $R_c$ from $G$ and $R$ as follows:

$$G = rG_c, \tag{18}$$
$$R = \frac{1}{r^3} R_c, \tag{19}$$

where $r$ is the radius of the acoustic tube. If $G$ and $R$ follow the theoretical equations (3) and (4), then $G_c$ and $R_c$ are

$$G_c = 2\pi \frac{\eta - 1}{\rho c^2} \sqrt{\frac{\lambda \omega_c}{2 c_p \rho}}, \tag{20}$$
$$R_c = \frac{2}{\pi} \sqrt{\frac{\omega_c \rho \mu}{2}}. \tag{21}$$

Eq. (20) and (21) encompass various physical properties of air; however, accurately measuring all these properties at the location where the acoustic tube is installed is challenging. Therefore, we directly compute the estimated values of $\hat{G}_c$ and $\hat{R}_c$ using PINNs, and calculate $\hat{G}$ and $\hat{R}$ as follows:

$$\hat{G} = r\hat{G}_c, \tag{22}$$
$$\hat{R} = \frac{1}{r^3} \hat{R}_c. \tag{23}$$

Therefore, for loss parameter identification, only two variables, $\hat{G}_c$ and $\hat{R}_c$, need to be identified using PINNs.

For the identification of $G$ and $R$, the data loss $L_M$ with respect to the measured sound pressure data at $x = l$ is defined as follows:

$$L_B = \frac{\lambda_M}{N_M} \sum_{i=1}^{N_M} (\hat{p}_i - \bar{p}_i)^2, \tag{24}$$

for $x_i = l$, where $N_M$ represents the number of collocation points for data loss, and $\bar{p}$ is the measured sound pressure data. The loss function $L_{inverse}$ for the entire system, used for parameter identification, is as follows:

$$L_{inverse} = L_E + L_B + L_C + L_P + L_B. \tag{25}$$

Finally, with $\hat{G}_c$ and $\hat{R}_c$ as learnable parameters, the loss parameter identification problem is formulated as

$$\min_{\theta,\hat{G}_c,\hat{R}_c} L_{inverse}\left(\theta,\hat{G}_c,\hat{R}_c\right) . \tag{26}$$

By minimizing the $L_{inverse}$ according to Eq. (26), the estimated values of the loss parameters $\hat{G}$ and $\hat{R}$ are calculated from the measured sound pressure data and the governing equations.

### 3.3 Implementation

We developed the PINNs using the Deep Learning Toolbox in MATLAB (MathWorks, USA). The "dlfeval" function was utilized to code a custom training loop, and the "dlaccelerate" function, along with the Parallel Computing Toolbox, was employed to expedite the computations. The training was conducted on a computer equipped with a Core i9-13900KS CPU (Intel, USA) and a GeForce RTX 4090 GPU (NVIDIA, USA). The system included 128 GB of main memory and 24 GB of video memory.

## 4. Validation of the proposed method

In this section, the validity of the proposed method is confirmed through simulations for loss parameter identification.

### 4.1 Forward analysis

Prior to parameter identification, the efficacy of PINNs is verified through forward analysis, assuming that the loss parameters $G$ and $R$ are known. The flow velocity waveform at $x = 0$ is provided as a boundary condition to PINNs, and the sound pressure waveform at $x = l$ is derived through forward analysis using Eq. (17) of PINNs.

The boundary condition at $x = 0$ comprises the Rosenberg wave, filtered by a low-pass filter with a cutoff frequency of 2 kHz, as shown in Fig. 2. The fundamental frequency of this waveform is 261.6 Hz (C4 in the musical scale), corresponding to $T = 3.82 \times 10^{-3}$ s. The physical properties employed in the simulation are detailed in Table 1. Following the methodology used in acoustic tube simulations in the time domain (Ishizaka and Flanagan, 1972), $\omega_c$ is considered constant. Regarding the structure of PINNs, the hidden layer consists of a fully-connected layer with 200 nodes; the number of fully connected (FC) blocks is set at 5. For further details on FC blocks, refer to our previous paper (Yokota, 2023). The number of collocation points $N_E$ is 5000, while $N_B$, $N_C$, and $N_P$ are each 1000.

Analytical results of the sound pressure waveform at $x = l$ are shown in Fig. 5. $x = l$ corresponds to the microphone position shown in Fig. 3 for parameter identification. The red line represents the results after training 100,000 epochs with PINNs, and the blue line indicates results from the finite difference method (FDM, CTCS scheme, $\Delta x = 10^{-3}$ m, $\Delta t = 0.5 \times 10^{-6}$ s). The alignment between the results from PINNs and FDM confirms that PINNs accurately learned the governing equations. The learning process for 100,000 epochs took 6,817 s using PINNs, highlighting the need for improved computational efficiency as a future objective.

Table 1 Analysis conditions.

| Parameter | Value |
|---|---|
| Air density $\rho$ | 1.20 kg/m$^3$ |
| Bulk modulus $K$ | 1.39×10$^5$ Pa |
| Speed of sound $c$ | 340 m/s |
| Viscosity coefficient $\mu$ | 19.0×10$^{-6}$ Pa·s |
| Heat-capacity ratio $\eta$ | 1.40 |
| Thermal conductivity $\lambda$ | 2.41×10$^{-2}$ W/(m·K) |
| Specific heat for constant pressure $c_p$ | 1.01 kJ/(kg·K) |
| Angular velocity for loss term $\omega_c$ | 1.64×10$^3$ rad/s |
| $l$, $d_1$, $d_2$ of tube | 0.1, 0.01, 0.02 m |

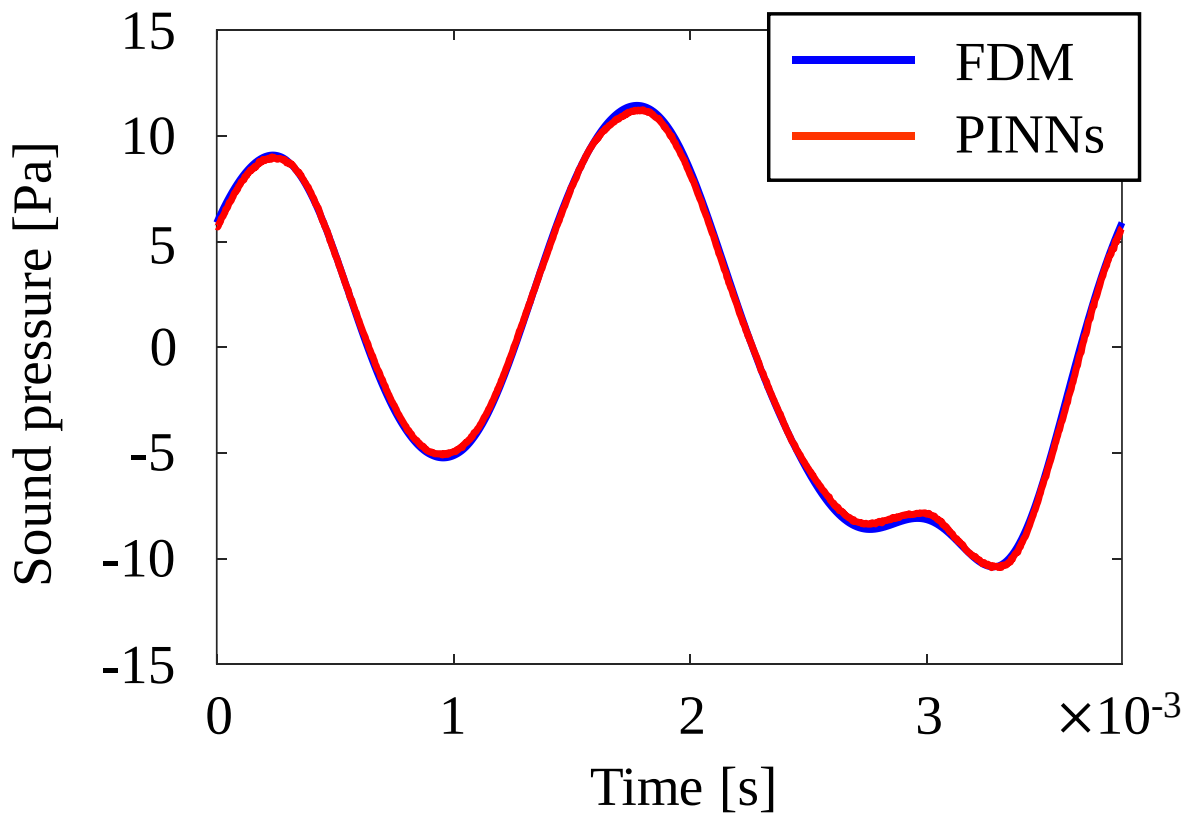

Fig. 5 Analyzed sound pressure waveform at $x = l$. The PINN results are in good agreement with those obtained by the finite difference method (FDM).

### 4.2 Validation of loss parameters identification

As described in Section 3.2, $G_c$ and $R_c$ are identified using the particle velocity waveform at $x = 0$ and the measured sound pressure waveform at $x = l$, assuming $G$ and $R$ are unknown. In this section, we conduct this parameter identification using PINNs described in Section 3 and evaluate its performance.

The physical properties employed in parameter identification are the same as those listed in Table 1. The particle velocity waveform at $x = 0$ is the Rosenberg wave shown in Fig. 2, consistent with the previous section. The sound pressure waveform at $x = l$, which is assumed to be measured using a microphone, is derived from the FDM as depicted in Fig. 5. The initial values of $G_c$ and $R_c$ are set with errors of +50% and -50% relative to the true values, respectively. The hyperparameters of the neural network remain consistent with those described in Section 4.1, and the loss parameters are identified by minimizing Eq. (26) with $\hat{G}_c$ and $\hat{R}_c$ as learnable parameters.

Table 2 presents the identification results of $G_c$ and $R_c$ after 100,000 epochs of training. Assuming that noise is introduced during the microphone measurement, the results are also displayed when 1% uncorrelated Gaussian noise is added to the sound pressure data $\bar{p}$. Regardless of the presence of noise, the identification results for $G_c$ after 100,000 epochs show an error margin of approximately 1–2%, compared with an initial error of 50%, indicating high accuracy in identification. Figure 6 illustrates the identification results for each epoch. It took 6,432 s to train for 100,000 epochs, and to stabilize the learning process in PINNs, the updating of $G_c$ and $R_c$ was halted for the first 10,000 epochs.

The identification results for $R_c$ show a relatively large discrepancy from the true value. This discrepancy arises because $G_c$ and $R_c$ have different influences on the changes in the sound pressure waveform at $x = l$ when their values are adjusted. Figure 7 displays the FDM simulation results of sound pressure changes at $x = l$ for varying values of $G_c$ and $R_c$. The gray lines are simulations using the loss parameters calculated from the properties in Table 1, which are utilized in parameter identification, and these loss values are denoted as $G_0$ and $R_0$. The red line represents the case where $R_c = 2R_0$, and the blue line where $G_c = 2G_0$. Evidently, changes in *Gc* exert a more pronounced impact on the shape of the sound pressure waveform than those in $R_c$. Thus, even if the value of $R_c$ changes, the sound pressure waveform at $x = l$ remains relatively unaffected. This observation suggests that the method, which identifies loss parameters from sound pressure data at $x = l$, did not yield high accuracy for $R_c$. Since PINNs approximate the solution of partial differential equations using the universal approximation theorem (Hornik et al., 1989) of neural networks, improvements to the structure and training methods of neural networks are considered necessary to enhance accuracy further.

Table 2 Identified $G$ and $R$ after training for 100,000 epochs.

| | $G_c$ [m/(Pa·s)] | $R_c$ [Pa·s/m] |
|---|---|---|
| True | 7.29×10$^{-5}$ | 8.73×10$^{-2}$ |
| Identified (no noise) | 7.40×10$^{-5}$ (1.6% error) | 5.03×10$^{-2}$ (42.4% error) |
| Identified (1% noise) | 7.40×10$^{-5}$ (1.6% error) | 4.43×10$^{-2}$ (49.2% error) |

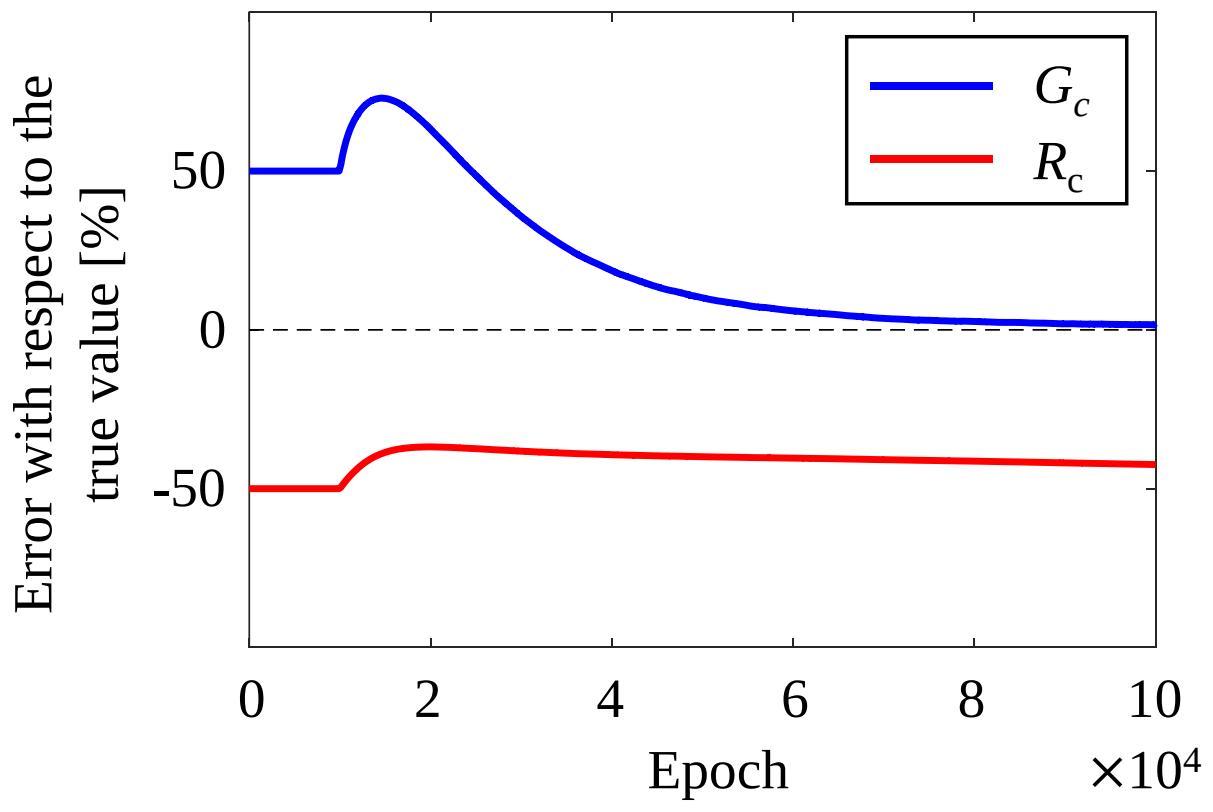

Fig. 6 Errors of the identified $G_c$ and $R_c$ for each epoch. $R_c$ error is relatively large, while the $G_c$ error approaches 0%.

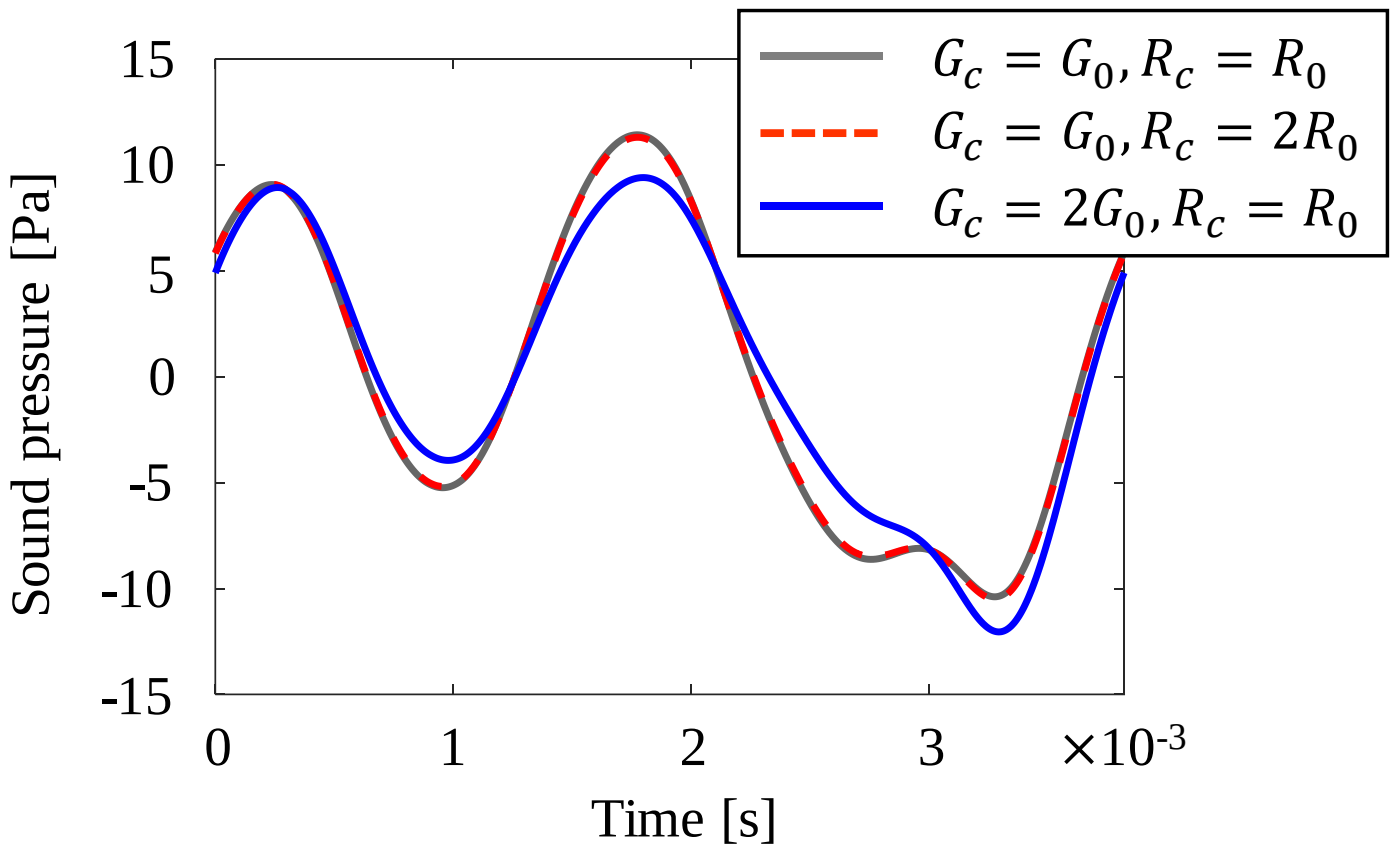

Fig. 7 FDM simulation results of sound pressure at $x = l$ for different loss parameters. Changes in $R_c$ have negligible effect on the waveform.

Despite the issues mentioned, changes in the value of $R_c$ have minimal impact on the sound pressure waveform at $x = l$. High-accuracy parameter identification is achieved for $G_c$, which significantly influences the acoustic phenomena addressed in this study. This highlights the broad applicability of PINNs for solving inverse problems related to acoustic tubes.

## 5. Conclusions

This study introduced a method for loss parameter identification in acoustic tubes using PINNs and evaluated the performance of PINNs, highlighting its potential for broad application in the inverse analysis of acoustic tubes. The loss parameters were treated as learnable variables, independent of the tube diameter, and the identification method was formulated as a neural network optimization problem using PINNs. By incorporating the particle velocity waveform at one end of the tube and the sound pressure waveform at the other end as loss functions in PINNs, we successfully identified the loss parameters $G$ and $R$ that satisfy the governing equations. Although the identification error was relatively large for the loss parameter that does not significantly influence the shape of the sound pressure waveform at $x = l$, the loss parameter that substantially affects the waveform could be identified with high accuracy.

The method developed in this study does not necessitate additional derivations such as transfer matrices and can be adapted to various acoustic problems by merely altering the loss function relative to the governing equations. Despite these advantages, we also identified a current limitation of PINNs for acoustic problems: there are instances where parameter identification lacks accuracy. In the future, we aim to enhance the accuracy and extend the applications of inverse analysis of acoustic tubes using PINNs by refining the algorithm, building on the identified advantages and limitations.

## Acknowledgements

This work is supported by JSPS KAKENHI Grant Number JP22K14447.